\documentclass[a4paper]{article}

\usepackage{amsmath}
\usepackage{amsfonts}
\usepackage{amssymb}
\usepackage{graphicx}
\DeclareGraphicsExtensions{.eps}

\begin{document}
\title{Stabilization of a Time Dependent Hamiltonian System}

\date{}

\author{Asher Yahalom$^{a,b}$ \& Natalia Puzanov$^a$   \\
$^a$Ariel University, Ariel 40700, Israel\\
$^b$Princeton University, Princeton, New Jersey 08543, USA
}

\maketitle

\bigskip

\begin{abstract}
In this paper we consider the unstable chaotic attractor of the Toda potential and stabilize it by a control in integral form.
In order to obtain stability results, we propose a special technique which is based on the idea of reduction of integro-differential
equations to  ordinary differential equations system.
\end{abstract}
\medskip

\medskip
\section {Introduction.\label{Intr}}
The presence of chaos in physical systems has been extensively demonstrated
and is very common.  In practice, however, it is often desired that chaos should be avoided and that the  performance of system
should be stabilized.  A review of various methods of controlling chaos is
presented in \cite{Andr}.

Various stabilization approaches are based on a delayed feedback control scheme  \cite{KPyragas}.
This scheme involves a control signal $u (t)$ generated by
the difference between the current state $x (t)$ and the previous state $x (t - \tau)$
 of the system.
 Examples of the use of this method can be found in electronic chaos oscillators \cite{Tamasev},
 magnetoelastic chaos \cite{Hiki}, lasers \cite{Biel}, the low-dimensional chaos arising from the nonlinear interaction between the  two different types of ionization waves \cite{Maus}, chemical systems \cite{Par}, \cite{Lekebusch},  plasma  \cite {Fukuyama}, \cite{Pierre},
Lorenz attractor \cite{Postlethwaite}.
Various modifications of this method have been proposed in \cite{Soc}, where
 an extended delayed feedback controller, using information about
many previous states of the system, was suggested.
More examples can be found in a review by M.A.F. Sanjuan and  C. Grebogi  \cite{Sanjuan}.

 Consider, for example, the unstable system of ordinary differential equations

\begin{equation*}
X^{\prime }(t)-A(t)X(t)=0,
\eqno (1)
\end{equation*}
where $A(t)$ is $n\times n$ matrix and $X(t)=col\left\{
x_{1}(t),...,x_{n}(t)\right\} .$ Adding a small external force $u(t)$ in the
right hand side of this system, we try to obtain various types of regular
behavior of the process $X(t).$ This force $u(t)$\ can be considered as a
control. It is clear that $u(t)$ should depend on the process $X(t).$ We thus obtain
the system:
\begin{equation*}
X^{\prime }(t)-A(t)X(t)+Ku(t)=0,
\eqno(2)
\end{equation*}
where $K$ is a corresponding $n\times n$ matrix.
Usually the control $u(t)$ depends on the values of the process $X(s)$ for $0 \leq s \leq t- \tau$,
where $\tau > 0$ is a time of a possible reaction, typical to real
systems. One of the simple ideas to achieve a stabilization is to choose the
control $u(t)$\ in the form $u(t)=X(t-\tau ).$\ The stabilization is
achieved only in the case when the system:
\begin{equation*}
X^{\prime }(t)-A(t)X(t)+KX(t-\tau )=0,
\eqno (3)
\end{equation*}
is stable. Describing existing approaches to the study of stability, we consider an autonomous systems i.e. with the case when all the coefficients
and delays in system (3) are constants.

Another approach to cope with delays is a reduction model approach,
also denoted the finite spectrum assignment technique, which originated in the works  of \cite{Artstein,Kwon,Manitius,Olbrot}. Recent developments of this concept are
discussed in \cite{Mazenc} (see also the references therein).  A development of this idea is to choose $u(t)$\ in the form of
the sum:
\begin{equation*}
u(t)=\sum_{i=1}^{m}K_{i}X(t-\tau _{i}).
\eqno (4)
\end{equation*}
It looks very natural to choose the control $u(t)$\ in the form:
\begin{equation*}
u(t)=\int_{0}^{t}k(t,s)X(s)ds.
\eqno (5)
\end{equation*}
in which all the history of the process $X(t)$ is taken into account. It will be convenient  for our stabilization purpose to choose:
\begin{equation*}
 k(t,s) = e^{-\beta (t-s)}.
\eqno (6)
\end{equation*}
There are two strong objections against the use of
the control in integral form. The first one is the following: we have ``to
remember'' all the values of a solution $X(s)$ for $s\in \lbrack 0,t]$ for
computing simulations. The second one is connected with the fact that
integral terms accumulate calculation errors. The method we suggest avoids these
objections. Using the exponential kernels, we reduce the study of
integro-differential system of the order $n$ to analysis of $(n+m)$-th order
system of ordinary differential equations \cite{Dom,Goltser}.

\textbf{Example}. Consider the following scalar equation
\begin{equation*}
x^{\prime }(t)=a_{1}x(t)
\eqno (7)
\end{equation*}
 $a_{1}$ is a constant. If $a_1>0,$ then  solution of equation (7): $x=0$ is unstable.
 Let us use the control $u(t)$ in the form (5), where $k(t,s)=e^{-\beta (t-s)}$ and consider the
integro-differential equation:
\begin{equation*}
 x^{\prime }(t)-a_1x(t)+\alpha u(t) = x^{\prime }(t)-a_1x(t)+\alpha\int_{0}^{t}  e^{-\beta
(t-s)}x(s)ds=0,\;t\in \lbrack 0,+\infty ).\eqno(8)
\end{equation*}
where $\alpha$ and $\beta$ are control parameters.

In accordance with the Leibnitz rule of differentiation under the sigh of an integral depending
on a parameter and limits of integration depend on the differentiation variable:
\begin{equation*}
\dfrac{d}{dy}\int_{a(y)}^{b(y)}f(x,y)dt = \int_{a(y)}^{b(y)}
\dfrac{\partial}{\partial y}f(x,y)dx + b^{\prime }(y)f(y,b(y)) - a^{\prime }(y)f(y,a(y))
\eqno(9)
\end{equation*}
we obtain:
\begin{equation*}
u^{\prime }(t) = -\beta\int_{0}^{t}e^{-\beta (t-s)}x(s)ds + e^{-\beta(t-t)} x(t) - 0 = -\beta u(t) +x(t)
\eqno(10)
\end{equation*}
Using Leibnitz rule, we can write the corresponding system in the form:
\begin{equation*}
\begin{cases}
x^{\prime }(t)=a_{1}x(t) - \alpha u(t), \\
u^{\prime }(t)  = -\beta u(t)  + x(t) =0,\,\,\,\,t\in \lbrack 0,+\infty ) \\
u(0) = 0
\end{cases}
\eqno(11)
\end{equation*}
Its characteristic equation is the following:
\begin{equation*}
\lambda ^{2}+(\beta - a_{1})\lambda +\alpha - \beta a_{1}=0.
\eqno(12)
\end{equation*}
The condition:
\begin{equation*}
\beta - a_{1} >0\ \ \text{and}\ \  \alpha - \beta a_{1}>0
\eqno (13)
\end{equation*}
is necessary and sufficient for the exponential stability of the system
(11). If condition (13) is fulfilled, then the solution $x(t)=0$ of the
equation (8) is exponentially stable.


\medskip
\section{Stabilization of Hamiltonian system by feedback control in integral form.}
\medskip

The state of a Hamiltonian system can be described by N generalized
momenta  $p \equiv (p_1, ... , p_N)$  and the same number  $N$  generalized coordinates
$q \equiv (q_1, ... , q_N)$. Here $N$ designates the number of a system's
degrees of freedom.  The evolution of $p$ and $q$ in time is determined by
the equations of motion:
\begin{equation*}
\dot{p_i} = -\frac {\partial{H}}{\partial{q_i}}; \  \  \dot{q_i} = \frac {\partial{H}}{\partial{p_i}};\  \  \left ( i = 1, ... ,N \right), \eqno(14)
\end{equation*}
which become concrete only with the Hamiltonian:
\begin{equation*}
H = \left (H,p,q,t\right).
 \eqno(15)
\end{equation*}
The Hamiltonian is given in $2N$ - dimensional phase space
$(p, q)$ and may also be an explicit function of time. Pairs of variables
$(p_i, q_i)$ are called canonically conjugate pairs and the equations (14)
are canonical equations.
We consider the two-dimensional motion may be defined by the Hamiltonian
\begin{equation*}
H = \frac {p_{x}^{2}}{2m} +  \frac {p_{y}^{2}}{2m} + V(x,y), \ \ m=1,
 \eqno(16)
\end{equation*}
where $V(x,y)$ is Toda lattice potential:
\begin{equation*}
V(x,y) = \frac {1}{2}(x^2+y^2) +x^2y -  \frac {1}{3}y^3 + \frac {3}{2}x^4 +  \frac {1}{2}y^4.
 \eqno(17)
\end{equation*}
 We rewrite system (14) in the form:
\begin{equation*}
\begin{cases}
\dot{x} = p_{x} \\
\dot{y} = p_{y} \\
\dot{p_x}=-\frac{\partial H}{\partial x} \\
\dot{p_y}=-\frac{\partial H}{\partial y}.%
\end{cases}%
 \eqno(18)
\end{equation*}%
or in more concrete form:
\begin{equation*}
\begin{cases}
\dot{x} = p_{x} \\
\dot{y} = p_{y} \\
\dot{p_x} = -x-2xy-6x^{3} \\
\dot{p_y} = -y-x^{2}+y^{2}-2y^{3}.%
\end{cases}%
 \eqno(19)
\end{equation*}
We represent this system in the time dependent explicit form:
\begin{equation*}
\begin{cases}
x^{\prime }(t)=p(t) \\
p^{\prime }(t)=-x(t)-2x(t)y(t)-6x^{3}(t) \\
y^{\prime }(t)=q(t) \\
q^{\prime }(t)=-y(t)-x^{2}(t)+y^{2}(t)-2y^{3}(t).%
\end{cases}%
 \eqno(20)
\end{equation*}
System (20) is autonomous and homogeneous.
Studying the stability of this system, we consider its linear approximation of the original nonlinear system at
the equilibrium point: $x=0, y=0, p=0, q=0$ according to Lyapunov's linearization method:
\begin{equation*}
\begin{cases}
x^{\prime }(t)=p(t) \\
p^{\prime }(t)=-x(t) \\
y^{\prime }(t)=q(t) \\
q^{\prime }(t)=-y(t).%
\end{cases}%
\eqno(21)
\end{equation*}
We rewrite system (21) in the vector form
\bigskip
\begin{equation*}
\left[
\begin{array}{c}
x^{\prime }(t ) \\
p^{\prime }(t ) \\
y^{\prime }(t ) \\
q^{\prime }(t )
\end{array}%
\right] =\left[
\begin{array}{cccc}
0 & 1 & 0 & 0 \\
-1 & 0 & 0 & 0 \\
 0 & 0 & 0 & 1 \\
 0 & 0 & -1 & 0%
\end{array}%
\right] \left[
\begin{array}{c}
x(t ) \\
p(t ) \\
y(t ) \\
q(t)%
\end{array}%
\right]  \eqno(22)
\end{equation*}
let us designate:
\begin{equation*}
A = \left[
\begin{array}{cccc}
0 & 1 & 0 & 0 \\
-1 & 0 & 0 & 0 \\
 0 & 0 & 0 & 1 \\
 0 & 0 & -1 & 0%
\end{array}%
\right] \eqno(23)
\end{equation*}
as the (constant) matrix of system (22). The determinant:
\bigskip
\begin{equation*}
det (A - \lambda I) =
\left \vert
\begin{array}{cccc}
 -\lambda & 1 & 0 & 0 \\
-1 &  -\lambda & 0 & 0 \\
 0 & 0 &  -\lambda & 1 \\
 0 & 0 & -1 &  -\lambda %
\end{array}%
\right \vert  \eqno(24)
\end{equation*}
leads to the characteristic equation $det(A-\lambda I)=0$ which will take the form:
\begin{equation*}
\lambda^4+2\lambda^2+1 = (\lambda^2+1)^2 = (\lambda - i)^2(\lambda + i)^2 = 0
\eqno(25)
\end{equation*}
and its characteristic roots  are
\begin{equation*}
\lambda_{1} = \lambda_{2} = i,  \, \, \, \, \lambda_{3} = \lambda_{4} = -i.
\eqno(26)
\end{equation*}
Now if:
\begin{equation*}
  det(A-\lambda I)=(\lambda - \lambda_1)^{m_{1}}(\lambda - \lambda_2)^{m_{2}} . . . (\lambda - \lambda_i)^{m_{i}} . . . (\lambda - \lambda_n)^{m_{n}}
\eqno(27)
\end{equation*}
then  elementary divisors are $(\lambda - \lambda_1)^{m_{1}}$ , . . . , $ (\lambda - \lambda_i)^{m_{i}}$ , . . .  .
Their number is the same as the number of the Jordan blocks of $A$  and the  elementary divisor $ (\lambda - \lambda_i)^{m_{i}}$ corresponds
to a Jordan blocks of oder $m_{i}$.  If ${m_{i}} = 1$ then  $(\lambda - \lambda_i)$ is a simple elementary divisor.
It follows that the Jordan form of (23)  consists of two blocks corresponding to each of the characteristic roots.

\bigskip

In accordance with the theorem, which was proved in the monograph \cite{Demid},
a linear homogeneous system with constant matrix A is stable if and only if all of its eigenvalues have nonpositive real parts, and the eigenvalues with zero real part may have only simple elementary divisors. That is, the corresponding Jordan cells are reduced to one element.

\bigskip
So the linear system (21) is chaotic behavior and hence also the nonlinear system (20)  is chaotic. The chaotic behaviour is demonstrated in Figure 1 where one can easily see than trajectories filling the area (see also \cite{Yahalom}).

\begin{center}
\scalebox{0.5}{\includegraphics*[0,0][580,200]{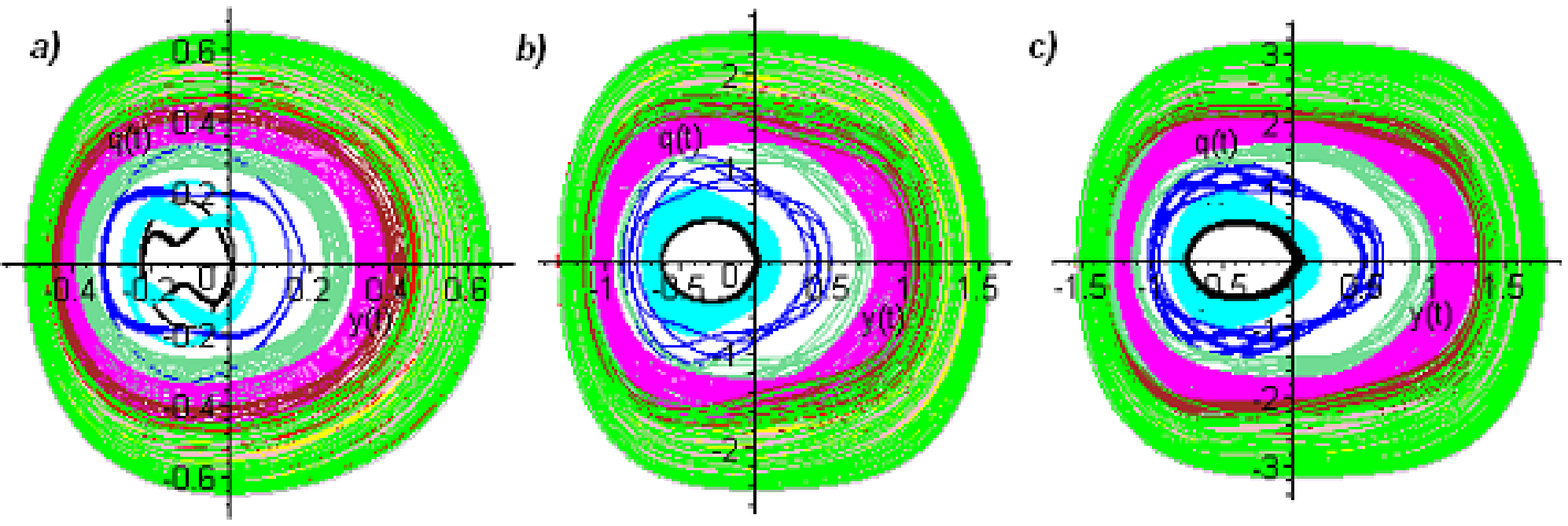}} \\
\small{\textbf{Figure 1}.} The phase portrait of the solution of system (20) in the ($y, q$) plane for\\a) $E = 0.214$, b) $E = 3.0$ and c) $E = 5.0$ with initial conditions: $x(0) = 0, \ y(0) = 0, \ p(0) = \sqrt {2E} \sin \left(\frac{a\pi}{2}\right), \  q(0) = \sqrt {2E} \cos \left(\frac{a\pi}{2}\right), $\\
For $ a = 0$ (gray), $a = 0.1$ (green), $a = 0.2$ (pink), $a = 0.3$ (yellow), $a = 0.4$ (red), $a = 0.5$  (brown), $a = 0.6$ (magenta), $a = 0.7$ (auqamarine), $a = 0.8$ (blue),\\
 $a = 0.9$ (cyan), $a = 1.0$ (black)
\end{center}
\vfil \eject

\section{Stabilization of Hamiltonian system by feedback control in integral form}

Let us use the control $u(t)$ in the form \cite{Dom2}:

\begin{equation*}
u(t)=\int\limits_{0}^{t}e^{-\beta (t-s)}q(s)ds,
\eqno(28)
\end{equation*}
in which all the history of the process q(t) is taken into account \cite{Dom}.
We apply stabilization to the second equation of system (20) by feedback delay control,
because for a mechanical system  control is normally applied to acceleration.
\begin{equation*}
\begin{cases}
x^{\prime }(t)=p(t) \\
p^{\prime }(t)=-x(t)-2x(t)y(t)-6x^{3}(t) -\alpha \int_{0}^{t}e^{-\beta (t-s)}p(s)ds \\
y^{\prime }(t)=q(t) \\
q^{\prime }(t)=-y(t)-x^{2}(t)+y^{2}(t)-2y^{3}(t)%
\end{cases}%
\eqno(29)
\end{equation*}%
where $\alpha $ and $\beta $ are control parameters.
We can rewrite the system (29) in a form of the system of ordinary differential equations according to Leibnitz rule (21):
\begin{equation*}
\begin{cases}
x^{\prime }(t)=p(t), \\
p^{\prime }(t)=-x(t)-2x(t)y(t)-6x^{3}(t) -\alpha u(t) \\
y^{\prime }(t)=q(t) \\
q^{\prime }(t)=-y(t)-x^{2}(t)+y^{2}(t)-2y^{3}(t) \\
u^{\prime }(t)=p(t)-\beta u(t).%
\end{cases}%
\eqno(30)
\end{equation*}%
System (30) is autonomous and homogeneous. Studying the stability of this system, we consider
its linear approximation of the original nonlinear system at the equilibrium point $x=0, y=0, p=0, q=0, u=0$:
\begin{equation*}
\begin{cases}
x^{\prime }(t)=p(t), \\
p^{\prime }(t)=-x(t)-\alpha u(t) \\
y^{\prime }(t)=q(t) \\
q^{\prime }(t)=-y(t) \\
u^{\prime }(t)=p(t)-\beta u(t)%
\end{cases}%
\eqno(31)
\end{equation*}
The constant matrix of the system (31) is:
\begin{equation*}
 A = \left[
\begin{array}{ccccc}
0 & 1 & 0 & 0 & 0\\
-1 & 0 & 0 & 0 & -\alpha \\
 0 & 0 & 0 & 1 & 0 \\
 0 & 0 & -1 & 0 &  0 \\
0  & 1 & 0 & 0 & -\beta%
\end{array}%
\right]  \eqno(32)
\end{equation*}
Which leads to the characteristic equation:
\begin{equation*}
det (A - \lambda I) =
\left \vert
\begin{array}{ccccc}
 -\lambda & 1 & 0 & 0 &  0 \\
-1 &  -\lambda & 0 & 0 & -\alpha\\
 0 & 0 &  -\lambda & 1 & 0\\
 0 & 0 & -1 &  -\lambda & 0 \\
 0 & 1 & 0 & 0 & -\lambda  -\beta%
\end{array}%
\right \vert  = 0
 \eqno(33)
\end{equation*}
The characteristic equation can be also written in the form:
\begin{equation*}
\begin{array}{cc}
- \left ( \lambda ^{5}+ \lambda ^{4}\beta+(\alpha
+2)\lambda ^{3} +2\beta \lambda ^{2}+(\alpha
+1)\lambda +\beta \right ) = \\
\\
- \left ( \lambda^3 +\lambda^2 \beta + \lambda (1+\alpha)+ \beta \right) \left (\lambda ^{2}+1 \right )=0%
\end{array}%
 \eqno(34)
\end{equation*}
and
\begin{equation*} 
 \lambda_{1} = \dfrac {C} {6} - \dfrac {2D} {C}  -  \dfrac {\beta} {3}; \, \, \, \, \, \,
\lambda_{2,3} = -\dfrac {C}{12} + \dfrac {D}{C} - \dfrac {\beta} {3}  \pm i  \sqrt{3} \left( \dfrac {C}{6} + \dfrac {2D}{C}\right )  ; \, \, \, \, \, \,
 \lambda_{4,5} = \pm i. %
 \eqno(35)
\end{equation*}
where
\begin{equation*}
\begin{array}{lcr}
A = 12 + 36\alpha +24 \beta^2 + 36\alpha^2 - 60\alpha\beta^2+ 12\beta^4+12\alpha^3 - 3\alpha^2\beta^2; \\
B = -72\beta + 36 \alpha\beta -8 \beta^3;
C = \left ( B + 12 \sqrt {A} \right )^ { {\dfrac {1} {3}}} ; \, \, \,  D = \dfrac {1} {3} \left (1 + \alpha -  \dfrac {\beta} {3} \right ).
\end{array}%
 \eqno(36)
\end{equation*}
For $\alpha > 0$,  $\beta > 0$   $\lambda_{1} $  and real parts of  $\lambda_{2}$  and  $\lambda_{3}$ are negative.
According to J. L. Massera's theorem, this is a stability indication for nonlinear ODE systems \cite{Demid}.
Massera's theorem states that given a nonlinear homogeneous system:
\begin{equation*}
\frac {dy}{dt} = A(t)y + f(t,y); \,\,  f(t,0) \equiv 0,
\eqno(37)
\end{equation*}
the  function $y$ is limited, If:

\textbf{1)}
\begin{equation*}
\|f(t,y)\| \leq \psi(t) \|y\|^m, \qquad (m>1)
\eqno(38)
\end{equation*}
where $\psi(t)$  is a positive function such that:
\begin{equation*}
\lim_{x\to\infty}\frac {1}{t}ln|\psi(t)| = 0
\eqno(39)
\end{equation*}

\textbf{2)} for Lyapunov characteristic exponents $a_1,  ... a_n$ of the linear approximation of a nonlinear system:
\begin{equation*}
\frac {dy}{dt} = A(t)y
\eqno(40)
\end{equation*}
the inequality:
\begin{equation*}
 max_{k} (a_k)  < -\displaystyle\frac {\kappa}{m-1}  \leq 0
\eqno(41)
\end{equation*}
 is fulfilled, where:
\begin{equation*}
 \kappa = \sum\limits_{k=1}^{n}a_k -  \lim_{x\to\infty}\frac {1}{t}\int_{0}^{t} tr \{A(t_1)\} dt_1
\eqno(42)
\end{equation*}
then the solution $y \equiv  0$ of a nonlinear system is asymptotically stable (Lyapunov) for $t\rightarrow \infty$.
In the neighborhood of $y=0$, which is the equilibrium position of the system (30),  nonlinear terms are smaller then linear ones.
System (30) is a system with constant coefficients (autonomous), therefore conditions (38) and (39)  are satisfied, i.e. condition \textbf{1)} fulfilled.

Since the characteristic exponents  $a_j (j=1, ... , n)$ of solutions of a linear system
\begin{equation*}
 \frac {dy}{dt} = Ay(t)
\eqno(43)
\end{equation*}
with a constant matrix $A$ are the real parts of the characteristic roots of the matrix $A$, i.e.
\begin{equation*}
  a_j = Re \lambda_j(A) \quad (j=1, ... , n)
\eqno(44)
\end{equation*}
 where $\lambda_j =\lambda_j(A)$ are roots of the equation $det(A-\lambda E) = 0$ \cite{Demid} then condition \textbf{2)} is fulfilled as
\begin{equation*}
max_{k} (a_k) = max(\lambda_{1}, Re\lambda_{2,3}) <0,
\eqno(45)
\end{equation*}
\begin{equation*}
 \sum\limits_{k=1}^{3}a_k =  \lambda_{1}+2Re\lambda_{2,3} = - \beta
\eqno(46)
\end{equation*}
and
\begin{equation*}
\lim_{x\to\infty}\frac {1}{t}\int_{0}^{t} tr \{A\} dt_1 =  - \beta.
\eqno(47)
\end{equation*}
Thus
\begin{equation*}
 \kappa = \sum\limits_{k=1}^{3}a_k -  \lim_{x\to\infty}\frac {1}{t}\int_{0}^{t} tr \{A\} dt_1 = 0
\eqno(48)
\end{equation*}
and inequality (41)  is fulfilled.
The stabilized solution is demonstrated in figure 2 where the trajectories are periodic and are not surface filling.
\begin{center}
\scalebox{0.5}{\includegraphics*[0,0][580,200]{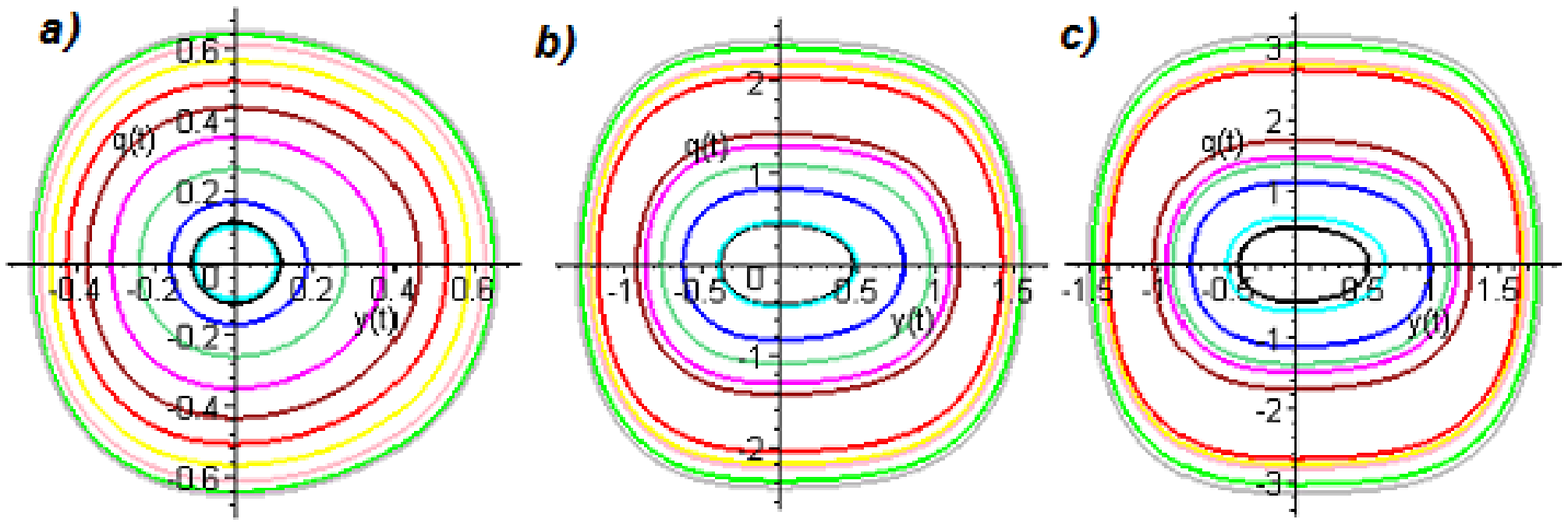}} \\
\small{\textbf{Figure 2}.} The phase portrait of the solution of system (19) in the ($y, q$) plane for\\  a) $E = 0.214$, b) $E = 3.0$ and c) $E = 5.0$ with initial conditions: $x(0) = 0, \ y(0) = 0, \ p(0) = \sqrt {2E} \sin \left(\frac{a\pi}{2}\right), \  q(0) = \sqrt {2E} \cos \left(\frac{a\pi}{2}\right), $\\$ \alpha = 2.0, \beta =1.0, \quad  40 \leq t \leq 300$ \\
For $ a = 0$ (gray), $a = 0.1$ (green), $a = 0.2$ (pink), $a = 0.3$ (yellow), $a = 0.4$ (red), $a = 0.5$  (brown), $a = 0.6$ (magenta), $a = 0.7$ (auqamarine), $a = 0.8$ (blue),\\
 $a = 0.9$ (cyan), $a = 1.0$ (black)
\end{center}
\vfil \eject
\section{Conclusion}

This model satisfies the rapid convergence of the process to a limit cycle with a small control function $u(t)$.
\begin{center}
\scalebox{0.6}{\includegraphics*[0,0][410,220]{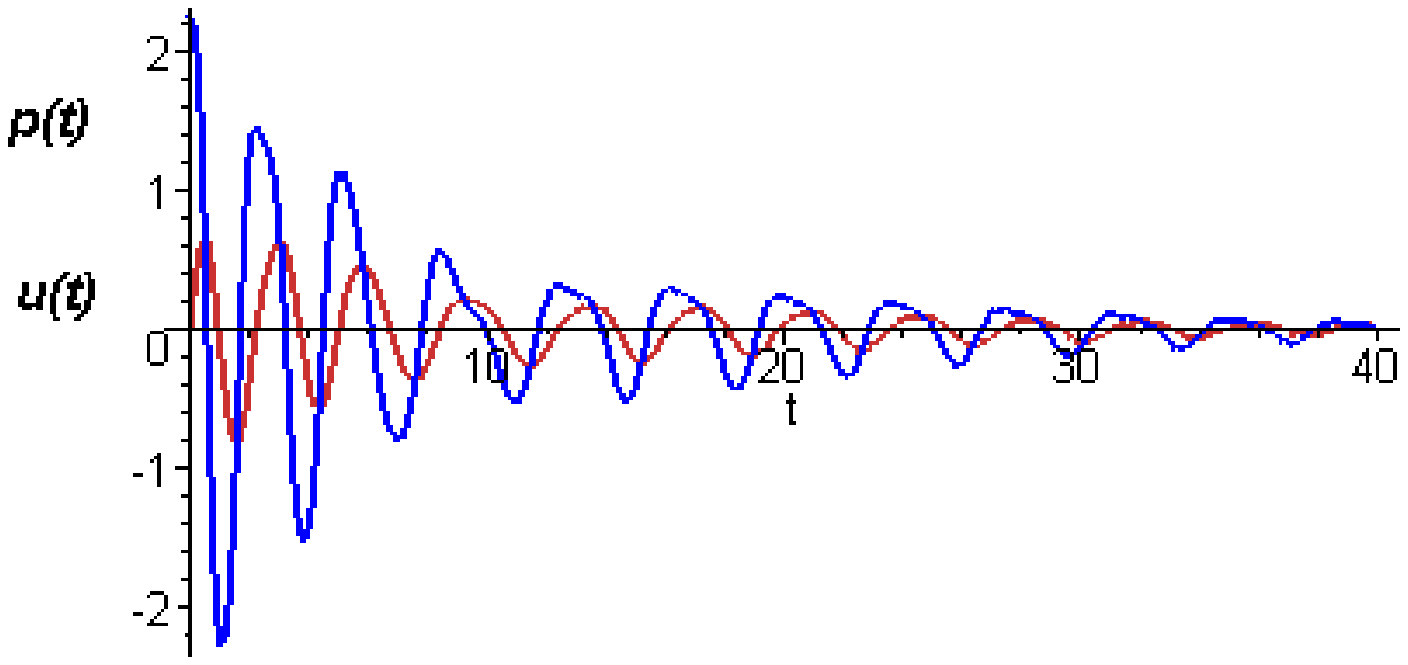}} \\
\small{\textbf{Figure 3}.} The control function $u(t)$ (red) as compared to the stabilized process \\ $p(t)$ (blue).
 $E=5.0, \alpha = 2.0, \beta =1.0, a=0.5$
\end{center}
This model differs from the methods described in publications of other
authors by the absence of an adjustable parameters and rough approximations in
determining of the control function.

It should be noted that the period of limit cycle obtained by
above-described method is not associated with the "period" of a chaotic
motion but is automatically obtained by the substitution of the values of
the control parameters $\alpha $ and $\beta $ to the nonlinear differential equations  system.

\end{document}